\documentclass[12pt]{article}
\usepackage{arxiv}
\usepackage{pythonhighlight}

\title{\texttt{BayesBlend}: Easy Model Blending using Pseudo-Bayesian Model Averaging, Stacking and Hierarchical Stacking in Python}
\author[12]{Nathaniel Haines\thanks{nathaniel.haines@ledgerinvesting.com}}
\author[12]{Conor Goold\thanks{conor@ledgerinvesting.com}}
\affil[1]{Ledger Investing, Inc.}
\affil[2]{\footnotesize\equalauth}
\date{\today}

\begin{document}
\maketitle

\begin{abstract}
    Averaging predictions from multiple competing inferential models 
frequently outperforms predictions from any single model,
providing that models are optimally weighted to maximize predictive performance.
This is particularly the case in so-called $\mathcal{M}$-open settings
where the true model is not in the set of candidate models, and
may be neither mathematically reifiable nor known precisely.
This practice of \textit{model averaging} has a rich history
in statistics and machine learning, and there are
currently a number of methods to estimate the weights for constructing
model-averaged predictive distributions. Nonetheless, there
are few existing software packages that can 
estimate model weights from the full variety of methods available,
and none that blend model predictions into a coherent predictive
distribution according to the estimated weights. In this paper, we introduce the
\texttt{BayesBlend} \textsf{Python} package,
which provides a user-friendly programming interface
to estimate weights and blend multiple (Bayesian) models' predictive distributions.
\texttt{BayesBlend} implements
pseudo-Bayesian model averaging, stacking and,
uniquely, hierarchical Bayesian stacking to estimate model weights.
We demonstrate the usage of \texttt{BayesBlend} with 
examples of insurance loss modeling.

\end{abstract}

\textit{Keywords}: ensemble learning, bayesian, model selection, open source

\section{Introduction}
Model averaging, part of ensemble learning \citep{murphy2012}, combines
predictions from multiple inferential statistical models rather than relying
on any one single model's predictions. Model averaging techniques have a rich history in 
statistics and machine learning \citep{bates1969,wolpert1992,breiman1996bagging,hoeting1999,hastie2009}
due to their improved performance over `winner-takes-all' approaches
that favor models with the highest rank-order performance.
While statisticians are familiar with estimating uncertainty or risk in
the parameters of their models, model structure is commonly
treated as known without error \citep{hodges1987,draper1995,hoeting1999}.
However, combining the predictive power of multiple plausible models 
typically leads to predictions far closer to the true data generating
process \citep{hoeting1999,draper1995,kaplan2021,yao2018stacking}.
This is particularly true in so-called $\mathcal{M}$-open settings \citep{bernardo2009,clyde2013,le2017},
where the true model, $M_{T}$, is neither known nor reifiable and, therefore,
not in the set of $K$ candidate
models being compared, $\mathcal{M} = \{M_{1}, M_{2}, ..., M_{K}\}$.
The central problem in model averaging is thus how
to estimate the weights, $\mathbf{w} = (w_1, ..., w_K)$,
to blend model outputs such that the resulting predictive distribution,
$M_{*}$, approximates the true data generating distribution from
$M_{T}$.

There are currently no off-the-shelf software implementations that allow quantitative 
researchers, statisticians, or data scientists to flexibly apply model averaging to their 
own model ensembles. We believe that this lack of software implementations plays a large part 
in why model averaging is not commonly observed in the applied statistics literature, which is 
surprising given its aforementioned utility for predictive modeling.

Here we introduce \texttt{BayesBlend}, which offers an easy programming interface in \textsf{Python} 
for Bayesian modeling workflows to $i$) estimate weights from multiple candidate models using a variety 
of published methods, and $ii$) blend model predictions into a coherent averaged distribution for 
downstream inference. We first introduce the relevant background literature on averaging Bayesian models 
necessary to understand the scope of \texttt{BayesBlend} in section \ref{section:background},
highlight existing software for averaging Bayesian models in section \ref{section:existing-software},
introduce \texttt{BayesBlend} formally in section \ref{section:bayesblend}, and provide
real-world examples of using \texttt{BayesBlend} from our professional domain of insurance loss
modeling in section \ref{section:examples}. It is our hope that \texttt{BayesBlend} and the associated
real-world examples we work through inspire researchers and practitioners alike to consider model 
averaging in their own work. 

\section{Background and scope}\label{section:background}
In applied Bayesian analysis, averaging Bayesian models has traditionally focused
on the technique of \textit{Bayesian model averaging} (BMA) \citep{raftery1997,hoeting1999,clyde1999,clyde2004}.
If Bayesian parameter estimation is the re-allocation of credibility across
possible parameter values, then
BMA is the re-allocation of credibility
across possible models \citep{kruschke2014,kruschke2011}. That is, BMA
estimates the posterior probabilities of each model, $M_{k}$, given
the data, $y$, and uses those \textit{posterior model probabilities} to
weight the differing model predictions.
The posterior model probabilities are defined formally using
Bayes' rule:

\begin{equation}
    p(M_{k} \mid y) = \frac{
        p(y \mid M_{k}) p(M_{k})
      }{
          \sum_{k=1}^{K} p(y \mid M_{k}) p(M_{k})
       } \propto 
        p(y \mid M_{k}) p(M_{k})
\end{equation}

where 

\begin{equation}
    \label{eq:marginal-likelihood}
    p(y \mid M_{k}) = \int p(y \mid \theta_{k}, M_{k}) p(\theta_{k} \mid M_{k}) d\theta_{k}
\end{equation}

is the marginal likelihood or prior predictive distribution 
of model $M_{k}$ with parameters 
$\theta_{k}$ \citep{raftery1997,hoeting1999},
$p(y \mid \theta_{k}, M_{k})$ 
is the likelihood density for model $M_{k}$, 
$p(\theta_{k} \mid M_{k})$ is the prior density for parameters
$\theta_{k}$ from model $M_{k}$,
and the denominator is a normalizing constant.
While the model probabilities can be estimated alongside each candidate model jointly
in a mixture modeling framework \citep{raftery1997,kruschke2014,keller2018}, it is more common
to estimate each candidate model separately and subsequently derive the posterior model
probabilities via estimation of the marginal likelihood and (often user-defined) prior model
probabilities, $p(M_{k})$, which are renormalized to sum to 1. 
Posterior model probabilities also have a close connection to the Bayes factor
(the ratio of marginal likelihoods) \citep{hinne2020}.
For instance, when $K=2$ and $p(M_1) = p(M_2) = \frac{1}{2}$,
then the posterior model probability of model 1 
is $p(M = 1 \mid y) = \frac{\mathrm{BF}}{1 + \mathrm{BF}}$,
where $p(M = 2 \mid y) = 1 - p(M = 1 \mid y)$. 

With the posterior model probabilities in hand, the model averaged posterior
predictive distribution for new data $\tilde{y}$ is defined as
as a linear mixture of models in $\mathcal{M}$:

\begin{align}
\begin{split}
    p(\tilde{y} \mid y) &= \sum_{k=1}^{K} p(\tilde{y} \mid y, M_{k}) p(M_{k} \mid y)\\
                        &= \sum_{k=1}^{K} \int p(\tilde{y} \mid \theta_{k}, M_{k}) p(\theta_{k} \mid M_{k}, y) p(M_{k} \mid y) d\theta_k
\end{split}
\end{align}

While BMA has been used extensively in a number of fields \citep{hinne2020},
it is frequently distinguished from other methods of ensemble learning 
in that it is ideally suited to $\mathcal{M}$-closed settings
where the true model, $\mathcal{M}_{T}$, is part of the candidate
model set \citep{minka2000,murphy2012}. Re-allocating credibility
across candidate models in $\mathcal{M}$ implictly assumes that all there is to
learn about the model uncertainty is in the candidate set of models and the training data. 
As the amount of data grows, $N \to \infty$, BMA will give 100\% weight to the model with 
the highest in-sample performance according to the lowest Kullback-Leibler 
divergence \citep{yao2018stacking}. 
Given that we may never truly know $\mathcal{M}_{T}$ in real applied modeling
problems, BMA may fail to provide weights that produce a blended predictive distribution
closest to the distribution of the true data generating process
in the harder $\mathcal{M}$-open settings.

One method of extending the space of candidate models under consideration
is to derive model weights from some estimate of out-of-sample
predictive performance. Determining model performance on new or future data
helps identify the level of generalization error in each model, and naturally
accounts for uncertainty in the space of possible models.
\cite{yao2018stacking} defined \textit{pseudo-Bayesian model averaging}
(pseudo-BMA) as the estimation of model weights by a renormalization of each candidate
model's expected log pointwise predictive densities (ELPD). ELPD estimates the expected
predictive performance of model $k$ on new data, $\tilde{y}$, and is, therefore, calculated
on hold-out test data or, more commonly, approximated by methods of cross-validation, like
leave-one-out (LOO) cross-validation, $\mathrm{elpd}_{\mathrm{loo}}$.
\cite{yao2018stacking} recommend using Pareto-smoothed importance sampling
(PSIS) estimates of leave-one-out (LOO) cross validation (PSIS-LOO) \citep{vehtari2017}
to estimate $\mathrm{elpd}_{\mathrm{loo}}$:

\begin{align}
\begin{split}
    \label{eq:elpd-loo}
    \mathrm{elpd}^{k}_{\mathrm{loo}} &= \sum_{i=1}^{N} \log p(y_{i} \mid y_{-i}, M_{k})\\
                                    &= \sum_{i=1}^{N} \log \int {
                                    p(y_{i} \mid \theta_{k}, M_{k}) p(\theta_{k} \mid M_{k}, y_{-i}) d\theta_{k}
                                    }\\
                                    &\approx \sum_{i=1}^{N} \log \frac{\sum_{s=1}^S r_{ik}^{s} p(y_{i} \mid \theta_{k}^{s}, M_{k})}{
                                    \sum_{s=1}^{s} r_{ik}^{s}}\\
                                    &= \widehat{\mathrm{elpd}}^{k}_{\mathrm{psis-loo}}
\end{split}
\end{align}

where $p(y_{i} \mid y_{-i})$ is the LOO predictive density of the $i$th data point given
data that does not include the $i$th data point, and
$p(y_{i} \mid \theta_{k}^{s}, M_{k})$ is the predictive density of 
the $i$th data point from the $s$th sample of the posterior distribution
of parameters $\theta_{k}$ and model $M_{k}$. To approximate the LOO predictive
density, a corrective importance sampling weight is applied to the vector
of (log) predictive densities, $r_{ik}$,
which is robustly estimated using PSIS \citep{vehtari2017}. 

The vector of ELPD values across models can then be transformed
using the softmax function into a set of weights:

\begin{equation}
    w_{k} = \frac{\exp(\widehat{\mathrm{elpd}}^{k}_{\mathrm{psis-loo}})}{
    \sum_{k=1}^{K} \exp({\widehat{\mathrm{elpd}}^{k}_{\mathrm{psis-loo}}})}
\end{equation}

Additionally, pseudo-BMA+ \citep{yao2018stacking} extends the normal pseudo-BMA weights above by
bootstrapping the log predictive densities $p(y_{i} \mid y_{-i})$
to further capture additional structural uncertainty in the candidate models. 

Outside of applied Bayesian analysis,
the machine learning literature has employed numerous approaches
to creating \textit{ensembles} of models that also emphasise out-of-sample
predictive accuracy. Techniques such as bagging \citep{breiman1996bagging}, 
majority voting \citep{murphy2012,xu1992}, and, most relevant here, 
\textit{stacking} \citep{wolpert1992,breiman1996stacking,tingwitten1997,murphy2012},
distinguish training a set of models on one data set while estimating the ensemble 
weights via estimators strictly on a hold-out data set.
The former models and data are sometimes referred to as
level-0 generalizers (i.e. models) and data \citep{wolpert1992,tingwitten1997,hastie2009},
while the latter are the level-1 models and data. 

Although model stacking is not new \citep{wolpert1992,breiman1996stacking},
it has relatively recently become a principal method of model averaging 
for Bayesian models \citep{clyde2013,yao2018stacking,yao2022stacking,yao2022stackingmix}.
While pseudo-BMA and pseudo-BMA+ collapse the PSIS-LOO log predictive densities
into $\widehat{\mathrm{elpd}}_{\mathrm{psis-loo}}$, stacking uses the PSIS-LOO
densities to find the the vector of weights that maximize the average log predictive
densities across models \citep{yao2022stacking}:

\begin{equation}
    \label{eq:stacking}
    \hat{\mathbf{w}} = \mathrm{arg} \max_{\mathbf{w}} \sum_{i}^{N} \log \sum_{k=1}^{K} w_{k} p(y_{i} \mid y_{-i}, M_{k})
\end{equation}

The stacking weights are most commonly subject, in applied Bayesian analyses,
to sum-to-one and non-negativity constraints (i.e. estimated as a unit simplex), 
but we note that this is not always the case \citep{clyde2013,le2017,breiman1996stacking}.
This optimization problem is a version of `complete-pooling' stacking, where stacking
weights are fixed with respect to the data index $i = (1, ..., N)$.
However, by re-writing equation \ref{eq:stacking} to reflect a more general 
inference problem, such as:

\begin{align}
\begin{split}
    p(y_i \mid y_{-i}) &= \sum_{k=1}^{K} w_{ik} p(y_{i} \mid y_{-i}, M_{k})\\
    w_{ik} &= \mathrm{softmax}(\mathbf{w}_{i1:K}^{*})\\
    \mathbf{w}_{i}^{*} &= f(\cdot)\\
\end{split}
\label{eq:bayes-stacking}
\end{align}

we can extend the model to no-pooling and partial-pooling
contexts, where unconstrained weights, $\mathbf{w}_{i}^{*}$, 
depend on the $i$th data index and can
be a function of arbitrary inputs, $f(\cdot)$, via a multinomial
regression using the softmax inverse-link function.
In particular, \cite{yao2022stacking} introduced hierarchical Bayesian stacking,
which can include continuous and discrete covariates as a form of 
\textit{feature-weighted}
stacking \citep{sill2009}. Consequently, a matrix of weights
$\mathbf{W}$ is returned. The model parameters, such as the covariates' coefficients,
can be partially-pooled to some global mean across candidate models. 
For instance, the following model partially-pools the coefficient $\beta$
for covariate $x$ across models:

\begin{align}
    \begin{split}
        p(y_i \mid y_{-i}) &= \sum_{k=1}^{K} w_{ik} p(y_{i} \mid y_{-i}, M_{k})\\
        w_{ik} &= \mathrm{softmax}(\mathbf{w}_{i1:K}^{*})\\
        w_{ik}^{*} &= \alpha_{k} + \beta_{k} x_{i}\\ 
        (\beta_{1}, ..., \beta_{K})' &\sim \mathrm{Normal}(\mu, \sigma)
    \end{split}
    \label{eq:hier-stacking}
\end{align}

where the coefficients $\beta = \beta_{1:K}$ across models have a hyperprior distribution
with parameters $\mu$ and $\sigma$. In practice, this model requires fixing one of the
model's weights to zero for identifiability.

While stacking does not need to be used to maximize the LOO log predictive
density, modern Bayesian model comparisons typically use the log score
as the canonical measure of model performance \citep{yao2018stacking,vehtari2017}.

\subsection{Scope}

\texttt{BayesBlend} implements pseudo-BMA, pseudo-BMA+, complete-pooling 
stacking using both optimization algorithms and full Bayesian inference, and hierarchical
stacking using full Bayesian inference. It does not perform 
tradiditional BMA, as there is already a range of existing software for implementing
BMA. At the time of writing, \texttt{BayesBlend} is the only software to implement
full Bayesian inference for stacking and hierarchical stacking. Additionally, a central goal
of \texttt{BayesBlend} is to make it easy for users to generate a blended or averaged
predictive distribution after estimating model weights, a step that is currently
missing from existing software implementations, as is explained in the following
section.

Note that we have developed \texttt{BayesBlend} with a focus on stacking of predictions
from Bayesian models, but in principle \texttt{BayesBlend} can be used to stack 
non-Bayesian models as well.

\section{Existing software}\label{section:existing-software}
In this section, we compare and contrast existing software implementations to \texttt{BayesBlend}
that allow for users to obtain model averaging weights using pseudo-BMA and stacking. We split 
the implementations into software packages and free-standing code. 
Although there are a number of implementations of model blending,
most do not perform blending on out-of-sample data, and none to our knowledge allow for users to 
train models on one set of data and then predict weights and perform blending on out-of-sample data.
Additionally, there are no software packages at the time of writing that implement hierarchial
Bayesian stacking.

\subsection{Software packages}

\texttt{loo} \citep{vehtari2017,vehtari2024} is an \textsf{R} package that enables users to compute efficient approximations of leave-one-out (LOO) cross-validation on Bayesian models via Pareto-smoothed importance sampling (PSIS-LOO). Although the primary focus is on PSIS-LOO, \texttt{loo} also contains methods for estimating pseudo-BMA, pseudo-BMA+ and stacking weights given the PSIS-LOO values. As of \texttt{loo==2.7.0}, all model weight estimation routines are implemented per the \texttt{loo\_model\_weights} method. Pseudo-BMA and pseudo-BMA+ weights are derived from transformations of the model LOO approximations, and stacking weights are derived per an optimization routine with the target in equation \ref{eq:stacking}. In either case, the output is a \texttt{data.frame} of models and their associated weights. \texttt{loo} does not: allow for Bayesian estimation of stacking weights, estimate hierarchical stacking weights, or perform blending of candidate models given estimated weights. 

\texttt{ArviZ} \citep{kumar2019} is a \textsf{Python} package focused on exploratory analysis of Bayesian models. As of \texttt{arviz==0.18.0}, the \texttt{arviz.compare} method both computes PSIS-LOO metrics and estimates pseudo-BMA, pseudo-BMA+ and stacking weights given these metrics, returning a \texttt{DataFrame} of models and their associated weights (along with information criteria and related metrics). The \texttt{ArviZ} model weight implementation follows almost exactly from the \texttt{loo} \textsf{R} package, and in many ways can be viewed as a \textsf{Python} implementation of \texttt{loo}. Similar to \texttt{loo}, \texttt{ArviZ} does not allow Bayesian estimation of weights, estimate hierarchical stacking weights, or perform blending. \texttt{BayesBlend} was designed to do all of these, working in tandem with \texttt{ArviZ} if used to obtain PSIS-LOO metrics (\texttt{loo} could also be used in conjunction with \texttt{BayesBlend}, but one would need to read the \texttt{loo} outputs into a \textsf{Python} environment first).

\subsection{Free-standing code}

In addition to the more comprehensive software packages described in the previous section, there are publications and blog posts including example code that can be adapted for user-specific applications. Specifically, hierarchical stacking \citep{yao2022stacking} is a relatively new technique that is not yet implemented in full-service software packages. The example code included in \cite{yao2022stacking} was particularly helpful in developing the Stan code for hierarhcical stacking in \texttt{BayesBlend}, which extends upon the example code in a few ways that we detail later. Similarly, the \texttt{NumPyro} \citep{bingham2019pyro} documentation has a \href{https://github.com/pyro-ppl/numpyro/blob/d7159b8161066127c1031448b7cd2c90b71d35b2/notebooks/source/bayesian_hierarchical_stacking.ipynb}{code example} of how to estimate pointwise model weights using hierarchical stacking and then use those weights to blend together posterior predictions from the models.

\section{\texttt{BayesBlend}}\label{section:bayesblend}
When using statistical software, users are often interested in inference, prediction, or some combination of the two. Those primarily interested in inference often desire software features that make it easier to build and interpret bespoke statistical models, whereas those interested primarily in prediction desire useability of software in real-world production settings. \texttt{BayesBlend} was designed with this latter production use in mind, allowing users to develop a blending model on one set of data and use it to generate both in- or out-of-sample predictions. 

\texttt{BayesBlend} can be installed from \texttt{PyPI} using \texttt{pip}:

\begin{python}
pip install bayesblend
\end{python}

\texttt{BayesBlend} relies on a working installation of \texttt{CmdStan} \citep{cmdstan2024}
for fitting the Bayesian stacking models.
We recommend installing \texttt{CmdStan} using \texttt{CmdStanPy} \citep{cmdstanpy2024}, which
is already a dependency of \texttt{BayesBlend}, in a \textsf{Python} session using:

\begin{python}
from cmdstanpy import install_cmdstan
install_cmdstan()
\end{python}

\texttt{BayesBlend} is compatible with the latest versions of \texttt{CmdStan} (\texttt{2.34.1}) 
and \texttt{CmdStanPy} (\texttt{1.2.2}) at the time of writing.

The core functionality of \texttt{BayesBlend} is divided into two classes:

\begin{itemize}
    \item \texttt{Draws}
    \item \texttt{BayesBlendModel}
\end{itemize}

\subsection{Draws}

The \texttt{Draws} dataclass is a ``smart'' container for the log posterior predictive densities and posterior predictions for a given substantive model. For illustration, we initialize a \texttt{Draws} object with some dummy posterior quantities: 

\begin{python}
import bayesblend as bb
import numpy as np

log_lik = np.array([
    [1, 2, 3, 4, 5, 6, 7, 8, 9], 
    [1, 2, 3, 4, 5, 6, 7, 8, 9]
])
post_pred = np.array([
    [1, 2, 3, 4, 5, 6, 7, 8, 9], 
    [1, 2, 3, 4, 5, 6, 7, 8, 9]
])

draws = bb.Draws(log_lik = log_lik, post_pred = post_pred)
\end{python}

\texttt{Draws} is able to handle arrays of any shape, but the first dimension should always index the posterior samples (only 2 samples in the dummy example). Additionally, the shape of \texttt{log\_lik} and \texttt{post\_pred} should always match if both are specified. In the example above, both \texttt{log\_lik} and \texttt{post\_pred} above are of dimension 2x9, indicating 2 posterior samples across 9 different datapoints. However, we could have equally well specified a 2x3x3 array for each if the underlying posterior arrays had more than 2 dimensions. 

Once initialized, \texttt{Draws} has convenience methods that are used to access information pertaining to the underlying posterior arrays and re-shape them as needed for later model weighting and blending routines: 

\begin{python}
>>> draws.n_samples
2

>>> draws.n_datapoints
9

>>> draws.shape
(2, 9)

# log pointwise density == log_mean_exp(log_lik)
>>> draws.lpd
array([1., 2., 3., 4., 5., 6., 7., 8., 9.])

>>> draws.log_lik_2d
array([
    [1, 2, 3, 4, 5, 6, 7, 8, 9],
    [1, 2, 3, 4, 5, 6, 7, 8, 9]
])

>>> draws.post_pred_2d
array([
    [1, 2, 3, 4, 5, 6, 7, 8, 9],
    [1, 2, 3, 4, 5, 6, 7, 8, 9]
])
\end{python}

The 2D representations above are what \texttt{BayesBlend} uses when fitting the pseudo-BMA or stacking models, but the \texttt{log\_lik} and \texttt{post\_pred} arrays always retain their original shape. In this example, the 2D representation is the same as the original. Further details on these methods are described later in the \texttt{BayesBlendModel} section. 

\subsubsection{Integrations}

We don't expect that users will be manually initializing \texttt{Draws} objects in most cases. Instead, we expect that users have already fitted a Bayesian model using their software of choice, and that \texttt{Draws} objects will be initialized based on outputs from this external software. Therefore, \texttt{Draws} objects have special class methods to extract the log posterior predictive densities and posterior predictions from external model objects. Currently, these include both \texttt{CmdStanPy} and \texttt{ArviZ}, although we will continually add more integrations into the future:

\begin{python}
# a model fit with cmdstanpy
cmdstan_fit = ...
cmdstan_draws = bb.Draws.from_cmdstanpy(
    cmdstan_fit,
    log_lik_name = "loglik",
    post_pred_name = "y_rep",
)

# an arviz InferenceData object
arviz_data = ...
arviz_draws = bb.Draws.from_cmdstanpy(
    arviz_data,
    log_lik_name = "log_lik",
    post_pred_name = "posterior_pred",
)
\end{python}

Above, the \texttt{log\_lik\_name} and \texttt{post\_pred\_name} arguments allow for users to extract the posterior arrays needed for \texttt{Draws} even if the log posterior predictive densities and posterior prediction array names differ from the defaults in \texttt{BayesBlend}. 

\subsection{BayesBlendModel}

The \texttt{BayesBlendModel} class is an abstract base class that is subclassed into the following averaging and stacking models: 

\begin{itemize}
    \item \texttt{MleStacking}
    \item \texttt{BayesStacking}
    \item \texttt{HierarchicalBayesStacking}
    \item \texttt{PseudoBma}
\end{itemize}

Each of these models estimates model weights differently, but they all share the same basic workflow. All \texttt{BayesBlendModel} models take a dictionary of \texttt{Draws} objects as input (one for each underlying substantive model of interest). The core functionality of each \texttt{BayesBlendModel} is then housed in the \texttt{.fit} and \texttt{.predict} methods. The \texttt{.fit} method fits the associated averaging/stacking model across the dictionary of \texttt{Draws} objects, and the latter returns a new \texttt{Draws} object that blends together the log posterior predictive densities and posterior prediction arrays across the substantive models given the estimated model weights. 

We illustrate usage of each averaging and stacking model above by way of example. The example here is intended to cover basic usage, and we relegate more advanced modeling strategies to later sections. Note that the code in each subsection below builds upon the same example. Therefore, for readers interested in following along, we recommend running the code in each section in sequence. 

To start, we simulate data from a Bernoulli distribution with a succuss probability that increases across trials:

\begin{python}
import numpy as np

rng = np.random.default_rng(1)
    
# generate data
N = 100
p = np.linspace(0.01, 1, N)
y = rng.binomial(1, p)

# split into training, test, and validation sets
y_train = y[0::3]
y_test = y[1::3]
y_valid = y[2::3]
\end{python}

Next, we need to fit the substantive models that will end up being blended together. Note that we use \texttt{CmdStanPy} for the substantive models due to personal preference, but we emphasize that \texttt{BayesBlend} is agnostic to where the log posterior predictive densities and posterior predictions come from. For this example, we will use the Bernoulli model Stan code below (saved locally as \texttt{bernoulli.stan}):

\begin{python}
data {
    int<lower=0> N_train;
    int<lower=0> N_test;
    int<lower=0> N_valid;
    real<lower=0> alpha_prior;
    real<lower=0> beta_prior;
    array[N_train] int<lower=0, upper=1> y_train;
    array[N_test] int<lower=0, upper=1> y_test;
    array[N_valid] int<lower=0, upper=1> y_valid;
} 
parameters {
    real<lower=0, upper=1> theta;
}
model {
    theta ~ beta(alpha_prior, beta_prior);
    y_train ~ bernoulli(theta);
}
generated quantities {
    vector[N_train] post_pred_train;
    vector[N_train] log_lik_train;
    vector[N_test] post_pred_test;
    vector[N_test] log_lik_test;
    vector[N_valid] post_pred_valid;
    vector[N_valid] log_lik_valid;
        
    for (n in 1:N_train){
        log_lik_train[n] = bernoulli_lpmf(y_train[n] | theta);
        post_pred_train[n] = bernoulli_rng(theta);
    }
    for (n in 1:N_test){
        log_lik_test[n] = bernoulli_lpmf(y_test[n] | theta);
        post_pred_test[n] = bernoulli_rng(theta);
    }
    for (n in 1:N_valid){
        log_lik_valid[n] = bernoulli_lpmf(y_valid[n] | theta);
        post_pred_valid[n] = bernoulli_rng(theta);
    }
}
\end{python}

Note that in the \texttt{bernoulli.stan} code, we are saving out the log posterior predictive densities and posterior predictions separately for the training, test, and validation datasets. Later, we will show how to either: (1) use the \texttt{log\_lik\_train} output to obtain approximate leave-one-out cross-validation (PSIS-LOO \cite{vehtari2017}) metrics as input for \texttt{BayesBlendModel} models, or (2) use the \texttt{log\_lik\_test} output directly as input for \texttt{BayesBlendModel} models. The validation set will then be used to assess stacking model performance.

We then fit the Bernoulli model with two different sets of priors--one set implying a low success probability and the other set implying a high success probability: 

\begin{python}
import bayesblend as bb
from cmdstanpy import CmdStanModel

model = CmdStanModel(stan_file="bernoulli.stan")

stan_data = {
    "N_train": len(y_train),
    "N_test": len(y_test),
    "N_valid": len(y_valid),
    "y_train": y_train,
    "y_test": y_test,
    "y_valid": y_valid,
}

# priors implying lower vs higher p 
priors_m1 = {
    "alpha_prior": 5,
    "beta_prior": 45,
}
priors_m2 = {
    "alpha_prior": 45,
    "beta_prior": 5,
}

# fit substantive models (dict for later use)
fits = {
    "low_p": model.sample(chains=4, data=stan_data | priors_m1, seed=1),
    "high_p": model.sample(chains=4, data=stan_data | priors_m2, seed=1),
}
\end{python}

With our substantive models fit, the following sections describe usage for each available \texttt{BayesBlendModel}. Note that \texttt{BayesBlendModel}'s all share the same workflow, so we have organized sections below to build upon each other. We recommend reading each section in sequence to fully understand all relevant usage patterns.  

\subsubsection{MleStacking}

The \texttt{MleStacking} class estimates model stacking weights using the optimization routine described by \cite{yao2018stacking}. Thefore, the stacking weights obtained per \texttt{MleStacking} are analagous to those obtained from both \texttt{loo} and \texttt{ArviZ} \citep{vehtari2024, kumar2019}. Continuing our Bernoulli example from above, we can fit the stacking model as follows: 

\begin{python}
# use the out of sample log likelihood directly
mle_stacking_fit = bb.MleStacking.from_cmdstanpy(
    fits, 
    log_lik_name="log_lik_test",
    post_pred_name="post_pred_test"
)
mle_stacking_fit.fit()
\end{python}

The \texttt{MleStacking} model uses \texttt{scipy} \citep{scipy2020} to optimize the stacking weights according to the log
of the objective function in equation \ref{eq:stacking}.
After fitting, the \texttt{mle\_stacking\_fit} has attributes for the stacking weights and convergence diagnostics: 

\begin{python}
# .model_info is the scipy OptimizeResult object
>>> mle_stacking_fit.model_info
message: Optimization terminated successfully
success: True
status: 0
    fun: 22.73730499217772
    x: [ 5.240e-01  4.760e-01]
    nit: 3
    jac: [-3.300e+01 -3.300e+01]
    nfev: 4
    njev: 3

>>> mle_stacking_fit.weights
{'low_p': array([[0.52398246]]), 'high_p': array([[0.47601754]])}
\end{python}

Given successful model convergence, we can be more confident in interpreting the model weights. In this case, we see above that more weight is given to the model with lower priors on $p$ as opposed to higher priors. 

We can also blend the underlying model draws by calling the \texttt{.predict} method: 

\begin{python}
# generate blended draws of test data
mle_blend_test = mle_stacking_fit.predict(seed=1)
\end{python}

Here, \texttt{mle\_blend\_test} is a \texttt{Draws} object where the \texttt{log\_lik\_test} and \texttt{post\_pred\_test} arrays from the underlying Bernoulli models are blended together. Note that despite these arrays being out-of-sample with respect to the underlying substantive models, they are in-sample for the stacking model. To assess the performance of the stacking model, we can use it to blend draws on the validation set, which has not been used to fit either the substantive models or the stacking model: 

\begin{python}
# make Draws objects with validation data
validation_draws = {
    model: bb.Draws.from_cmdstanpy(
        fit, "log_lik_valid", "post_pred_valid"
    ) for model, fit in fits.items()
}
# blend validation Draws
mle_blend_valid = mle_stacking_fit.predict(validation_draws, seed=1)
\end{python}

We can then use the \texttt{.lpd} property to assess the relative performance of the stacking model on the test data (which it was trained on) vs validation data. We expect that validation performance will be slightly worse overall given it is out-of-sample:  

\begin{python}
# taking mean across pointwise LPD because of different N's
>>> mle_blend_test.lpd.mean()
-0.689042870360223

>>> mle_blend_valid.lpd.mean()
-0.6990944721829563
\end{python}

We see that the test blend indeed has a higher expected LPD. 

\subsubsection{BayesStacking}

The \texttt{BayesStacking} class is analagous to \texttt{MleStacking}, except it uses MCMC sampling (per \texttt{CmdStanPy}) to estimate model weights as opposed to an optimization routine. The use of Bayesian weight estimation allows for us to set more informed priors on weights, and in general it tends to produce weights that are regularized more toward uniform weighting compared to \texttt{MleStacking}. Usage of the \texttt{BayesStacking} model is almost identical to \texttt{MleStacking} apart from the ability to assign priors on weights:

\begin{python}
# Dirichlet prior on weights (defaults to uniform)
priors = {"w_prior": [1,1]}

bayes_stacking_fit = bb.BayesStacking.from_cmdstanpy(
    fits, 
    log_lik_name="log_lik_test",
    post_pred_name="post_pred_test",
    priors=priors,
    seed=1,
)
bayes_stacking_fit.fit()
\end{python}

After fitting, the \texttt{bayes\_stacking\_fit} object can also be inspected for weights and convergence information: 

\begin{python}
# .model_info is the CmdStanMCMC object, which can be summarized
>>> bayes_stacking_fit.model_info.summary()
         Mean   MCSE StdDev    5
lp__   -24.74  0.024  0.86 -26.51 -24.39 -24.12  1217.0   8818.9  0.999
w[1]     0.51  0.004  0.17   0.22   0.52   0.79  1384.7  10034.6  1.001
w[2]     0.48  0.004  0.17   0.20   0.47   0.77  1384.7  10034.6  1.001
ll[1]   -0.62  0.004  0.15  -0.90  -0.60  -0.39  1389.3  10067.8  1.001
ll[2]   -0.62  0.004  0.15  -0.90  -0.60  -0.39  1389.3  10067.8  1.001
ll[3]   -0.62  0.004  0.15  -0.90  -0.60  -0.39  1389.3  10067.8  1.001

>>> bayes_stacking_fit.weights
{'low_p': array([[0.51987123]]), 'high_p': array([[0.48012877]])}
\end{python}

Note that the \texttt{BayesStacking} weight values are very similar to the \texttt{MleStacking} weights from before. Convergence is also successful, but if there are convergence issues, we can change the \texttt{Stan} sampler behavior with the \texttt{cmdstan\_control} argument: 

\begin{python}
bayes_stacking_conv_fit = bb.BayesStacking.from_cmdstanpy(
    fits, 
    log_lik_name="log_lik_test",
    post_pred_name="post_pred_test",
    priors=priors,
    cmdstan_control={"adapt_delta": .95},
    seed=1,
)
bayes_stacking_conv_fit.fit()
\end{python}

Blending for \texttt{BayesStacking} models works the same as in \texttt{MleStacking}:

\begin{python}
# generate blended draws of test data
bayes_blend_test = bayes_stacking_fit.predict(seed=1)

# blend validation Draws (defined in MleStacking section)
bayes_blend_valid = bayes_stacking_fit.predict(validation_draws, seed=1)
\end{python}

\subsubsection{HierarchicalBayesStacking}

The \texttt{HierarchicalBayesStacking} model is the most flexible model in \texttt{BayesBlend}. ``Hierarchical'' refers to the internal structure of the model, which allows for users to estimate stacking weights conditional on covariates (per a multiple regression on the unconstrained weights; see equation \ref{eq:hier-stacking} and \cite{yao2022stacking}) while optionally pooling information across covariate coefficients. Therefore, hierarchical stacking is ideal for situations where we expect model weights to vary across some known predictor. 

In our example, recall that the ``true'' generative model is a Bernoulli model with increasing success probability across trials. If we think of each trial as a distinct unit in time, we can create a time covariate for the hierarhcical stacking model to better capture the generative process: 

\begin{python}
# create a "time" covariate for y
x = np.linspace(1, 100, N)

# same splits used for y
x_train = x[0::3]
x_test = x[1::3]
x_valid = x[2::3]
\end{python}

We can then use this covariate when fitting the hierarchical stacking model: 

\begin{python}
hier_stacking_fit = bb.HierarchicalBayesStacking.from_cmdstanpy(
    fits, 
    log_lik_name="log_lik_test",
    post_pred_name="post_pred_test",
    continuous_covariates={"time": x_test},
    continuous_covariates_transform="standardize", # default
    partial_pooling=False, # default
    seed=1,
)
hier_stacking_fit.fit()
\end{python}

The \texttt{continuous\_covariate\_transform} argument allows for users to choose from a variety of different transformations to apply to continuous covariates. By default, continuous covariates are standardized (i.e. $\bar{x} = \frac{x - \text{mean}(x)}{2\text{sd}(x)}$). Other options include \texttt{"identity"} and \texttt{"relu"} (rectified linear unit). In general, standardization works easily in practice because the default priors are on a scale that works well with standardized variables. When using other transforms, we caution users to set priors accordingly. The priors used for fitting can be inspected per the \texttt{.priors} attribute once the model is initialized and fit.

When \texttt{partial\_pooling=False}, pooling parameters across covariate coefficients are not freely estimated. Therefore, any pooling done across covariate coefficeints is done solely by the priors assigned to the coefficients (which default to standard normal distributions). 

When \texttt{partial\_pooling=True}, pooling parameters are freely estimated. The model assumes a three-level structure with a global coefficient mean at the top level, unique offsets for discrete and continuous coefficeints in the middle level, and then unique offsets for each individual covariate coefficient at the bottom level. Therefore, we only recommend setting \texttt{partial\_pooling=True} when there are at least 3 covariate coefficients or more in the model. Note that discrete covariates are automatically dummy-coded, so a single discrete covariate with 3 levels will meet this criteria. For more details, refer to the \texttt{HierarchicalBayesStacking} documentation. 

With the model fitted, convergence can be checked in the same way as with the \texttt{BayesStacking} model above. Unlike the \texttt{BayesStacking} model, because hierarhcical stacking estimates model weights conditional on covariates, we obtain an individual model weight for each datapoint: 

\begin{python}
# (note output is rounded for readability)
>>> hier_stacking_fit.weights
{
    'low_p': array([[
        0.79, 0.78, 0.77, 0.76, 0.74, 0.73, 0.71, 0.69, 0.68, 0.66, 0.64,
        0.62, 0.6 , 0.58, 0.56, 0.54, 0.52, 0.49, 0.47, 0.45, 0.43, 0.41,
        0.39, 0.37, 0.35, 0.33, 0.32, 0.3 , 0.28, 0.27, 0.25, 0.24, 0.23
    ]]), 
    'high_p': array([[
        0.21, 0.22, 0.23, 0.24, 0.26, 0.27, 0.29, 0.31, 0.32, 0.34, 0.36,
        0.38, 0.4 , 0.42, 0.44, 0.46, 0.48, 0.51, 0.53, 0.55, 0.57, 0.59,
        0.61, 0.63, 0.65, 0.67, 0.68, 0.7 , 0.72, 0.73, 0.75, 0.76, 0.77
    ]])
}
\end{python}

There is a clear pattern in the weights such that the model with higher probability priors on $p$ gets more weight across time. The true data generating process has an increasing Bernoulli success probability across time, and the hierarhcical stacking model is able to capture this trend. In fact, we can look at the estimated effect of time by inspecting the model coefficients: 

\begin{python}
>>> hier_stacking_fit.coefficients
{'alpha': array([[0.06943556]]), 'beta_cont': array([[[-1.76384032]]])}
\end{python}

Here, \texttt{alpha} is the model intercept, and \texttt{beta\_cont} is the $\beta$ weight estimated for the time covariate. The intercept indicates the average weight of the reference model when the (standardized) time covariate is 0, and the $\beta$ weight indicates the change in weight with a 2SD unit change in time (recall that standardized covariates take the form $\bar{x} = \frac{x - \text{mean}(x)}{2\text{sd}(x)}$). The reference model is always the first model in the dictionary used to to fit the \texttt{BayesBlendModel}, which is the \texttt{low\_p} model in our case. Therefore, the coefficients suggest a higher average weight on the \texttt{low\_p} model that decreases as a function of time. The higher average weight on the \texttt{low\_p} model is consistent with the weights derived from the \texttt{MleStacking} and \texttt{BayesStacking} models, both of which had a slight preference for the \texttt{low\_p} model. 

To better understand the model coefficients, recall that the regression model is on the unconstrained weights. Weights are normalized per a softmax function, so we can map \texttt{alpha} to the implied average model weight per a softmax transformation: 

\begin{python}
>>> np.exp(np.array([0.06943556, 0])) / sum(np.exp(np.array([0.06943556, 0])))
array([0.51735192, 0.48264808])
\end{python}

The weights above reflect average model weights for the \texttt{low\_p} and \texttt{low\_h} models, respectively, which are in strong agreement with the corresponding weights per the \texttt{MleStacking} and \texttt{BayesStacking} models.

Blending model draws with \texttt{HierarchicalBayesStacking} models is similar to other models, although the use of covariates adds complexity. For blending the data used to train the stacking model, the workflow is identical to other models: 

\begin{python}
# generate blended draws of test data
hier_blend_test = hier_stacking_fit.predict(seed=1)
\end{python}

However, to blend the validation data, we need to specify the covariates associated with the validation data: 

\begin{python}
# blend validation Draws (defined in MleStacking section)
hier_blend_valid = hier_stacking_fit.predict(
    validation_draws, 
    continuous_covariates={"time": x_valid},
    seed=1,
)
\end{python}

Note that the \texttt{HierarchicalBayesStacking.predict} method will transform the new continuous covariates with the same values used (if any) when transforming the covariates used to train the model. In our case, the mean and SD of the \texttt{x\_test} covariate are used to scale the \texttt{x\_valid} covariate before prediction. You can access these values through the \texttt{covariate\_info} attribute: 

\begin{python}
>>> hier_stacking_fit.covariate_info
{'time': {'mean': 50.0, '2sd': 57.1314274283428, 'median': 50.0}}
\end{python}

\subsubsection{PseudoBma}

The \texttt{PseudoBma} class estimates model weights using the pseudo Bayesian model averaging procedures (pseudo-BMA as well as pseudo-BMA+) outlined in \cite{yao2018stacking}. Therefore, weights estimated per \texttt{PseudoBma} are analagous to those obtained from both \texttt{loo} and \texttt{ArviZ} \citep{vehtari2024, kumar2019}. \texttt{PseudoBma} weights can be estimated as follows: 

\begin{python}
pbma_fit = bb.PseudoBma.from_cmdstanpy(
    fits, 
    log_lik_name="log_lik_test",
    post_pred_name="post_pred_test",
    bootstrap=False, 
)
pbma_fit.fit()
\end{python}

To obtain so-called pseudo-BMA+ weights (where weights are regularized per the Bayesian bootstrap; \cite{yao2018stacking}), we can set \texttt{bootstrap=True}: 

\begin{python}
pbma_plus_fit = bb.PseudoBma.from_cmdstanpy(
    fits, 
    log_lik_name="log_lik_test",
    post_pred_name="post_pred_test",
    bootstrap=True, # default
    seed=1,
)
pbma_plus_fit.fit()
\end{python}

In both cases, the underlying model used to estimate the weights relies on simple transformations of the log pointwise density (i.e. the \texttt{BayesBlendModel.lpd} property). Therefore, fitting the \texttt{PseudoBma} does not involve MCMC sampling or optimization (just bootstrapping in the case of pseudo-BMA+). 

Blending for \texttt{PseudoBma} models works the same as for other models:

\begin{python}
# generate blended draws of test data
pbma_blend_test = pbma_fit.predict(seed=1)
pbma_plus_blend_test = pbma_plus_fit.predict(seed=1)

# blend validation Draws (defined in MleStacking section)
pbma_blend_valid = pbma_fit.predict(validation_draws, seed=1)
pbma_plus_blend_valid = pbma_plus_fit.predict(validation_draws, seed=1)
\end{python}

\subsection{Other usage patterns}

\subsubsection{PSIS-LOO for model training}

In low-data settings, it may not be feasible to construct reasonably sized training, test, and validation datasets. Even in cases where we do have enough data, it may be computationally burdensome to obtain out-of-sample log likelihood estimates (e.g., through proper leave-one-out cross-validation). In such cases, PSIS-LOO can be used as direct input to \texttt{BayesBlend} models. To do so, we will continue from our ongoing example: 

\begin{python}
import arviz as az

# use ArviZ to estimate PSIS-LOO
az_data = {
    model: az.from_cmdstanpy(
        fit, 
        log_likelihood="log_lik_train",
        posterior_predictive="post_pred_train"
    ) for model, fit in fits.items() 
}
psis_loo = {
    model: az.loo(data).loo_i.values 
    for model, data in az_data.items()
}

mle_stacking_loo_fit = bb.MleStacking.from_lpd(
    lpd=psis_loo,
    post_pred={
        model: data.posterior_predictive["post_pred_train"]
        for model, data in az_data.items()
    }
)
mle_stacking_loo_fit.fit()
\end{python}

Above, we use \texttt{ArviZ} to obtain approximate PSIS-LOO values for each datapoint in the training data and then use these values directly to fit the stacking model using the \texttt{.from\_lpd} class method. The model then behaves the same as in previous examples. 

\subsubsection{Comparing blended models}

Once blended \texttt{Draws} objects have been obtained for a given model (or set of models), we can easily compare models by comparing the expected log pointwise predictive density (ELPD) across models. For example, the following code computes ELPD for each of the blended validation set predictions generated throughout the above examples: 

\begin{python}
# including non-blended candidate models for reference
candidate_models = {
    model: bb.Draws.from_cmdstanpy(
        fit,
        log_lik_name="log_lik_valid",
        post_pred_name="post_pred_valid"
    ) for model, fit in fits.items()
}

validation_blends = candidate_models | {
    "mle": mle_blend_valid,
    "bayes": bayes_blend_valid,
    "hierbayes": hier_blend_valid,
    "pbma": pbma_blend_valid,
    "pbma_plus": pbma_plus_blend_valid,
}
    
elpd_blends = {
    model: sum(draws.lpd) 
    for model, draws in validation_blends.items()
}
\end{python}

Now, we inspect the results: 

\begin{python}
# (output rounded for readability)
>>> elpd_blends
{
    'low_p': -29.37, 
    'high_p': -25.5, 
    'mle': -23.07, 
    'bayes': -23.06, 
    'hierbayes': -20.38, 
    'pbma': -23.37, 
    'pbma_plus': -23.1
}
\end{python}

Higher ELPD indicates better model performance, so these results suggest that the hierarchical stacking model produces better blended predictions compared to other models--an expected outcome given that the hierarchical stacking model is the only model that can account for the changing nature of the Bernoulli success probability over time. It is also notable that all the blended models perform better than either candidate model in isolation.

\section{Real world examples}\label{section:examples}
Here we use \texttt{BayesBlend} in two real-world examples that are often encountered in actuarial 
science and practice. The first involves modeling how insurance losses develop over time, which 
illustrates blending when the data are assumed to follow multiple different parametric curves. The 
second involves forecasting insurance losses, which illustrates blending when the data are assumed to 
follow multiple different time-series dynamics. 

For both examples, we use the data provided by \cite{meyers2015stochastic}, which contains historical 
loss data for 50 insurance programs. We touch on details of the dataset where relevant in each section 
below. We refer interested readers to our GitHub repository for scripts used to reproduce both examples: 
\url{https://github.com/LedgerInvesting/stacking-paper-2024}.

\subsection{Loss development}
\subsubsection{Background}

Insurance data is often organized into a so-called ``triangle" format, which can be 
thought of as a table with rows indexing a particular time period, and columns indexing 
the (cumulative or incremental) losses within each time period when looking back from future time periods:

\begin{table}[htbp]
    \centering
    \caption{Loss Insurance Triangle}
    \label{tab:triangle}
    \begin{tabular}{cccccc}
      \hline
      & \multicolumn{5}{c}{Development Lag} \\
      \cline{2-6}
      Accident Year & $DL_1$ & $DL_2$ & $DL_3$ & $\cdots$ & $DL_n$ \\
      \hline
      $AY_1$ & $L_{11}$ & $L_{12}$ & $L_{13}$ & $\cdots$ & $L_{1n}$ \\
      $AY_2$ & $L_{21}$ & $L_{22}$ & $L_{23}$ & $\cdots$ & $L_{2n}$ \\
      $AY_3$ & $L_{31}$ & $L_{32}$ & $L_{33}$ & $\cdots$ & $L_{3n}$ \\
      $\vdots$ & $\vdots$ & $\vdots$ & $\vdots$ & $\ddots$ & $\vdots$ \\
      $AY_m$ & $L_{m1}$ & $L_{m2}$ & $L_{m3}$ & $\cdots$ & $L_{mn}$ \\
      \hline
    \end{tabular}
\end{table}

In Table \ref{tab:triangle}, $AY_m$ indicates the accident year that a loss occurs in, 
and $DL_n$ indicates the number of years after the given accident year 
in which we are viewing losses. Values in each cell then indicate our 
view on the loss for a given accident year at a given development lag. 
For example, if $AY_1$ is 2010, $L_{11}$ is the loss for accident year 
2010 as we know it in 2011, $L_{12}$ is the loss for accident year 2010 
as we know it in 2012, and so on. Within each accident year, we expect 
losses to grow across development lags until they reach some asymptote, 
resulting in the \textit{ultimate loss} for a given accident year ($UL_m$). 
Note that insurance programs have many different types of loss, and the 
dataset we are working with in this example has both paid and incurred losses. 
Paid losses are losses that the insurer has paid out to policy holders, and 
incurred losses are those that the insurer has estimated will need to be
paid out for a given period of time. For our example, we will use paid losses. 

Taken alone, the ultimate loss is not a very useful for decision-making. 
Instead, we are typically interested in the \textit{ultimate loss ratio} 
for each accident year--the ratio of the ultimate loss to the premium 
earned for the given accident year ($ULR_m = UL_m / EP_m$). $ULR_m < 1$ 
then indicates that more premium is earned than losses incurred for a 
given accident year $m$. Generally, we expect that insurance programs 
will incur ultimate loss ratios between .5 and 1, although there is a 
great deal of fluctuation both within and across insurance programs. 

The practice of predicting the ultimate loss for each accident year is 
referred to as \textit{loss development}. Loss development is necessary 
to understand where we expect loss ratios for recent accident years to 
end up in the future. To illustrate the issue, refer to Table \ref{tab:triangle-dev}: 

\begin{table}[htbp]
    \centering
    \caption{Loss Development}
    \label{tab:triangle-dev}
    \begin{tabular}{ccccccccccc}
      \hline
      & \multicolumn{10}{c}{Development Lag} \\
      \cline{2-11}
      Accident Year & 1 & 2 & 3 & 4 & 5 & 6 & 7 & 8 & 9 & 10 \\
      \hline
      2014 & 0.15 & 0.26 & 0.34 & 0.40 & 0.44 & 0.48 & 0.50 & 0.52 & 0.53 & 0.54 \\
      2015 & 0.07 & 0.12 & 0.16 & 0.19 & 0.21 & 0.22 & 0.23 & 0.24 & 0.24 & -- \\
      2016 & 0.17 & 0.30 & 0.40 & 0.47 & 0.52 & 0.56 & 0.59 & 0.62 & -- & -- \\
      2017 & 0.18 & 0.32 & 0.43 & 0.52 & 0.59 & 0.64 & 0.68 & -- & -- & -- \\
      2018 & 0.16 & 0.30 & 0.41 & 0.50 & 0.57 & 0.63 & -- & -- & -- & -- \\
      2019 & 0.22 & 0.39 & 0.51 & 0.60 & 0.67 & -- & -- & -- & -- & -- \\
      2020 & 0.25 & 0.45 & 0.61 & 0.75 & -- & -- & -- & -- & -- & -- \\
      2021 & 0.13 & 0.22 & 0.28 & -- & -- & -- & -- & -- & -- & -- \\
      2022 & 0.16 & 0.26 & -- & -- & -- & -- & -- & -- & -- & -- \\
      2023 & 0.16 & -- & -- & -- & -- & -- & -- & -- & -- & -- \\
      \hline
    \end{tabular}
\end{table}

Table \ref{tab:triangle-dev} is constructed to mimic the information we would have on an 
insurance program's cumulative loss ratios as-of January, 2024. In 
2024, we can look back and see how our view on the loss ratio for 
accident year 2014 has developed for each year since then. We see 
that it has more-or-less stabilized, and we can be rather certain 
that the $ULR_{2014} \approx 0.54$. However, for accident year 2023, 
we only have 1 year of development available. For accident years 
between 2014-2023, we having varying amounts of missing information 
(creating the triangle shape that insurance triangles are named after). 
In fact, the right-edge diagonal is our latest view on the loss ratio 
for each accident year, and it is clear that many years are still 
developing. Loss development is concerned with using such historic 
loss triangles to predict the ultimate loss ratio for each accident year. 

\subsubsection{Modeling loss development}

Traditionally, models of loss development follow the form: 

\begin{align}
  \begin{split}
    LR_{mn} &\sim f(\mu_{mn}, \sigma^{2}_n)\\
    \mu_{mn} &= LR_{m(n-1)} \cdot ATA_{n-1} \\
  \end{split} \quad \forall n \in (2, ..., N)
  \end{align}

where $LR_{mn}$ is the loss ratio for accident year $m$ at 
developmemnt lag $n$, $f(.)$ is distribution that can be 
parameterized by a mean and variance, and $ATA_{n-1}$ is the 
\textit{age-to-age factor} for development lag $n-1$. In general, 
we expect that $\lim_{n \to \infty} ATA_{n-1} = 1$, capturing the 
idea that losses within each accident year will reach an 
asymptote as development lags increase. The number of development 
lags required to reach stability varies across lines of business 
and other program-specific factors.

For our purposes, we assume that age-to-age factors follow an 
exponential curve that decays to 1 as development lags increase. 
Along with the decay of age-to-age factors, we assume that the 
outcome variance $\sigma^2$ decreases as a function of development:

\begin{align}
\begin{split}
  ATA_{n-1} &= 1 + \exp(\beta_0 + \beta_1 DL_{m(n-1)}) \\
  \sigma^{2}_n &= \exp(\sigma_0 - \sigma_1 DL_{m(n-1)})^2 \\
\end{split}
\end{align}

Note that parametric curves on age-to-age factors are most often 
used by actuaries when modeling the tail-end of the loss development 
process, whereas non-parametric models are used to capture early 
development \citep{mack1993distribution}. For this example, we 
will use the tail model for the entire loss development process. 

Cumulative loss ratios are both positive by construction and can theoretically 
reach values much greater than 1 if large, unanticipated losses are 
incurred. Therefore, for the outcome distribution $f(.)$ in all the models
in this section and in the forecasting example below, we use 
the gamma distribution due to its positive domain and right skew. 
We present the models parameterizing the gamma
distribution by its mean, $\mu$, and variance, $\sigma^2$, which are converted to the
shape, $\alpha$, and rate/inverse scale, $\beta$, parameterization before model estimation
using $\alpha = \mu^2 / \sigma^2$ and $\beta = \mu / \sigma^2$, respectively.

Lastly, as we explain in the next section, we have loss triangles 
for multiple different insurance programs from the same line of 
business. Therefore, we fit the loss development model hierarchically with
a partial pooling structure across $\beta_0$ and $\beta_1$. Altogether, 
the full hierarchical exponential loss development model is defined 
as follows: 

\begin{align}
\begin{split}
    LR_{imn} &\sim \mathrm{Gamma} \left(\mu_{imn}, \sigma^{2}_n\right)\\
  \mu_{imn} &= LR_{im(n-1)} \cdot ATA_{i(n-1)}\\
  ATA_{i(n-1)} &= 1 + \exp(\beta_{0i} - \beta_{1i} DL_{im(n-1)})\\
  \sigma^{2}_n &= \exp(\sigma_0 - \sigma_1 DL_{im(n-1)})^2 \\
  \beta_{0i} &\sim \mathrm{Normal}(\beta_{0\mu}, \beta_{0\sigma})\\
  \beta_{1i} &\sim \mathrm{Lognormal}(\beta_{1\mu}, \beta_{1\sigma})\\
  \sigma_0 &\sim \mathrm{Normal}(-4, 1)\\
  \sigma_1 &\sim \mathrm{Normal}^+(.5, .25)\\
  \beta_{0\mu} &\sim \mathrm{Normal}(0, .5)\\
  \beta_{0\sigma} &\sim \mathrm{Normal}^+(0, .5)\\
  \beta_{1\mu} &\sim \mathrm{Normal}(-1, 1)\\
  \beta_{1\sigma} &\sim \mathrm{Normal}^+(0, 1)\\
\end{split} \quad \forall n \in (2, ..., N)
\end{align}

In practice, we use non-centered parameterizations for the hierarchical 
parameters to make sampling more efficient. We refer readers to our 
online code repository where they can access the Stan model code for 
more implementation details if interested.

\subsubsection{Candidate models}

The dataset provided by \cite{meyers2015stochastic} includes loss 
triangles for 50 insurance programs within multiple different lines 
of business. The dataset spans accident years from 1988 to 1997, 
with 10 years of development for each accident year. For our purposes, 
we use the private passenger auto data because auto losses tend to 
develop rather quickly, meaning that the 10 years of development in 
the dataset is sufficient enough for each accident year to reach a 
stable, near ultimate loss ratio. 

Before fitting the model, we divided each of the the 50 10x10 loss 
ratio triangles into training, testing, and validation datasets. The 
training set consisted of development years 1 through 7, the test set 
development years 8 through 9, and the validation set was only development 
year 10 (see Table \ref{tab:triangle-splits}). We used the training set to fit 
candidate models, the test set to estimate out-of-sample log posterior 
predictive densities for each of the candidate models (used to fit the 
\texttt{BayesBlend} models), and the validation set to compare the candidate 
and blending models. 

\begin{table}[htbp]
  \begin{threeparttable}
    \centering
    \caption{Train, Test, Validation Scheme}
    \label{tab:triangle-splits}
    \begin{tabular}{ccccccccccc}
      \hline
      & \multicolumn{10}{c}{Development Lag} \\
      \cline{2-11}
      Accident Year & 1 & 2 & 3 & 4 & 5 & 6 & 7 & 8 & 9 & 10 \\
      \hline
      1988 & Train & Train & Train & Train & Train & Train & Train & Test & Test & Val \\
      1989 & Train & Train & Train & Train & Train & Train & Train & Test & Test & Val \\
      1990 & Train & Train & Train & Train & Train & Train & Train & Test & Test & Val \\
      1991 & Train & Train & Train & Train & Train & Train & Train & Test & Test & Val \\
      1992 & Train & Train & Train & Train & Train & Train & Train & Test & Test & Val \\
      1993 & Train & Train & Train & Train & Train & Train & Train & Test & Test & Val \\
      1994 & Train & Train & Train & Train & Train & Train & Train & Test & Test & Val \\
      1995 & Train & Train & Train & Train & Train & Train & Train & Test & Test & Val \\
      1996 & Train & Train & Train & Train & Train & Train & Train & Test & Test & Val \\
      1997 & Train & Train & Train & Train & Train & Train & Train & Test & Test & Val \\
      \hline
    \end{tabular}
    \begin{tablenotes}
      \small
      \item Note that we treated test and validation data as unknown 
      quantities. Therefore, if $LR_{im(n-1)}$ was not in the ``known'' 
      training data, then posterior predictions $\widehat{LR_{im(n-1)}}$ 
      were used to propagate uncertainty forward across development. 
    \end{tablenotes}
  \end{threeparttable}
\end{table}

For our candidate models, as opposed to fitting different variations of 
development models, we instead fit the exponential model described above 
four separate times to different sections of the training data:

\begin{itemize}
  \item Exp 1+: fit to all development years
  \item Exp 2+: fit to development years 2+ 
  \item Exp 3+: fit to development years 3+
  \item Exp 4+: fit to development years 4+
\end{itemize}

These models mimic a common situation encountered in actuarial practice 
where tail development models are not fit to the entire development history, 
but instead only to later development years. Typically, actuaries would 
select such a cutoff by hand, but blending allows for us to let the data 
decide the optimal cutoff through weighting each model based on its 
performance on future hold-out data. 

\subsubsection{Blending models}

We used the 1000 out-of-sample log posterior density estimates (20 per program 
with 50 programs) to train each of the 5 available models in \texttt{BayesBlend}.
For the hierarchical stacking model, we included the development and accident
years as discrete covariates, allowing the model to put more or less weight 
on candidate models as a function of both general industry trends (for which accident 
year is a proxy) and how far along the program is in development (per development 
year). 

\begin{figure}
  \centering
  \includegraphics[scale=.3]{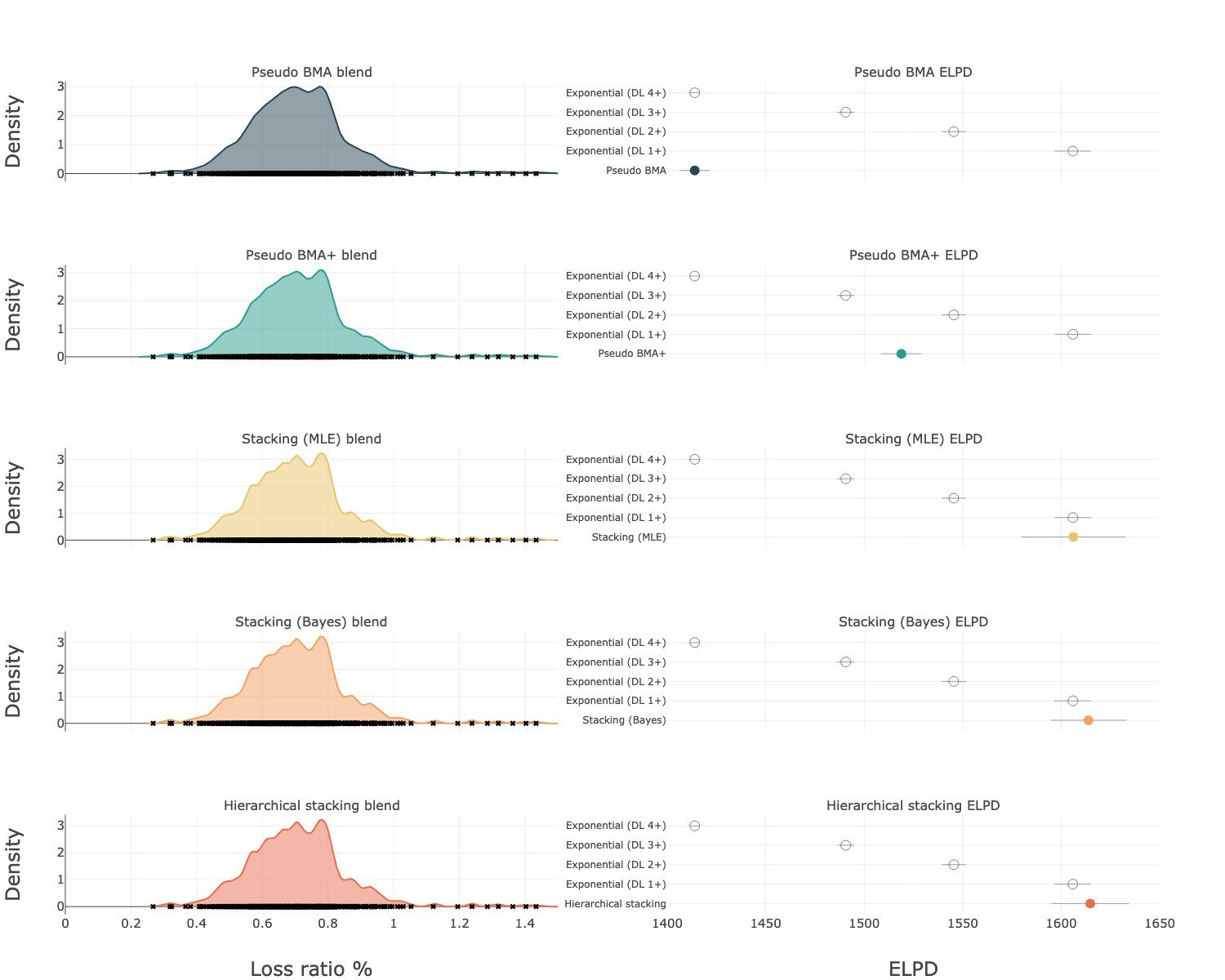}
  \caption{
      Blended posterior distributions and ELPD values from the
      deveopment example. In the first column, black crosses on 
      the blended posterior densities indicate the real data. 
      In the second column, ELPD point estimates and 1 standard 
      deviation ranges are shown by open and filled circles, denoting 
      candidate model and blended ELPD estimates, respectively, and 
      the horizontal line segments. 
  }
  \label{fig:development-scores}
\end{figure}

The blended posterior distributions and the (validation) ELPDs for each of the
four models and blended distributions is shown in Figure \ref{fig:development-scores}. 
The blended ELPD scores indicate that the Bayesian stacking models perform best,
with hierarchical stacking performing slightly better than the non-hierarhcical 
variant. Interestingly, MLE stacking produced blended ELPD scores 
that are practically indistinguishable from the best-fitting candidate model 
(the Exp 1+ model), and pseudo-BMA performed worse than the best candidate model. 
Upon further inspection, we found that the test set (used to train the blending 
models) had an outlier LPD point that was worse for the Exp 1+ model. Because 
pseudo-BMA estimates weights based on a simple transformation of the sum of LPDs
across datapoints, the outlier dominated the weight estimation. The stacking
models were not as sensitive to this strong outlier point. This illusrates a key
benefit of stacking--because it is model based (and informed by priors in the 
case of Bayesian stacking), it is more robust to outlier performance metrics 
compared to pseudo-BMA. 

\begin{figure}
  \centering
  \includegraphics[scale=.25]{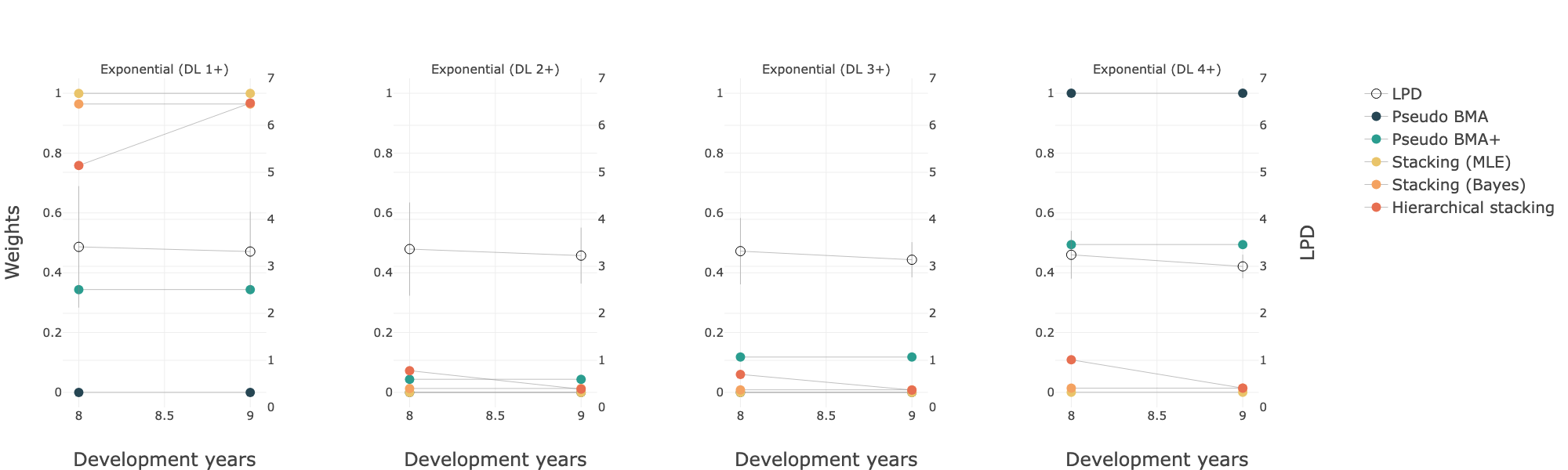}
  \caption{
      Model weights and log posterior predictive density (LPD) values from
      the loss development example. Model weights are constant across development 
      years for every model except the hierarchical stacking model, which 
      included development year as a discrete covariate. The left y-axis 
      illustrates the weight values for each of the blending models. The 
      right y-axis denotes the mean (open circles) and 1 standard deviation
      across programs (vertical line segments) LPD values.
  }
  \label{fig:development-weights}
\end{figure}

The model weights in Figure \ref{fig:development-weights} show that the Exp 1+ 
model is given most of the weight by both Bayesian stacking models, followed by a 
relatively uniform weight across other candidate models. The MLE 
stacking model places almost all the weight on Exp 1+ model, leaving little 
weight for other candidate models. The pseudo-BMA model gives almost all weight 
to Exp 4+, which as noted was the worse-performing model on the validation set. 
The pseudo-BMA+ model, however, gives a more even weighting between Exp 1+ and 
Exp 4+, which is expected due to the regularizing effect of the Bayesian bootstrap
procedure underlying the model. When looking at weights as a function of the 
development year, we see that the hierarchical stacking model gives more 
uniform weights across all candidate models at development year 8, then places most 
of the weight on the Exp 1+ model at development year 9. This illustrates the
ability of hierarchical stacking to assign better weights based on covariate values, 
which in this case leads to the slight performance improvement shown in Figure 
\ref{fig:development-scores}. 

Overall, these results suggest that stacking (particularly Bayesian variants) may 
offer a more data-driven, practical alternative to the traditional actuarial practice
of selecting which development years to use when training parameteric development models.
They also demonstrate the usefulness of hierarchical stacking in bringing external 
knowledge into a stacking model to produce better blended predictions.

\subsection{Forecasting}
Our second example uses the same data set as the development model example,
but instead focuses on the task of forecasting future accident period's
developed loss ratios. Once developed loss ratios are obtained, predictions
of future accident period's loss ratios can inform expected performance
of the book of business, which is of use to both business management
practices and as well as necessary for the  construction of cashflow patterns
important for capital release schedules in insurance-linked securities dynamics.

\begin{figure}
    \centering
    \includegraphics[scale=0.17]{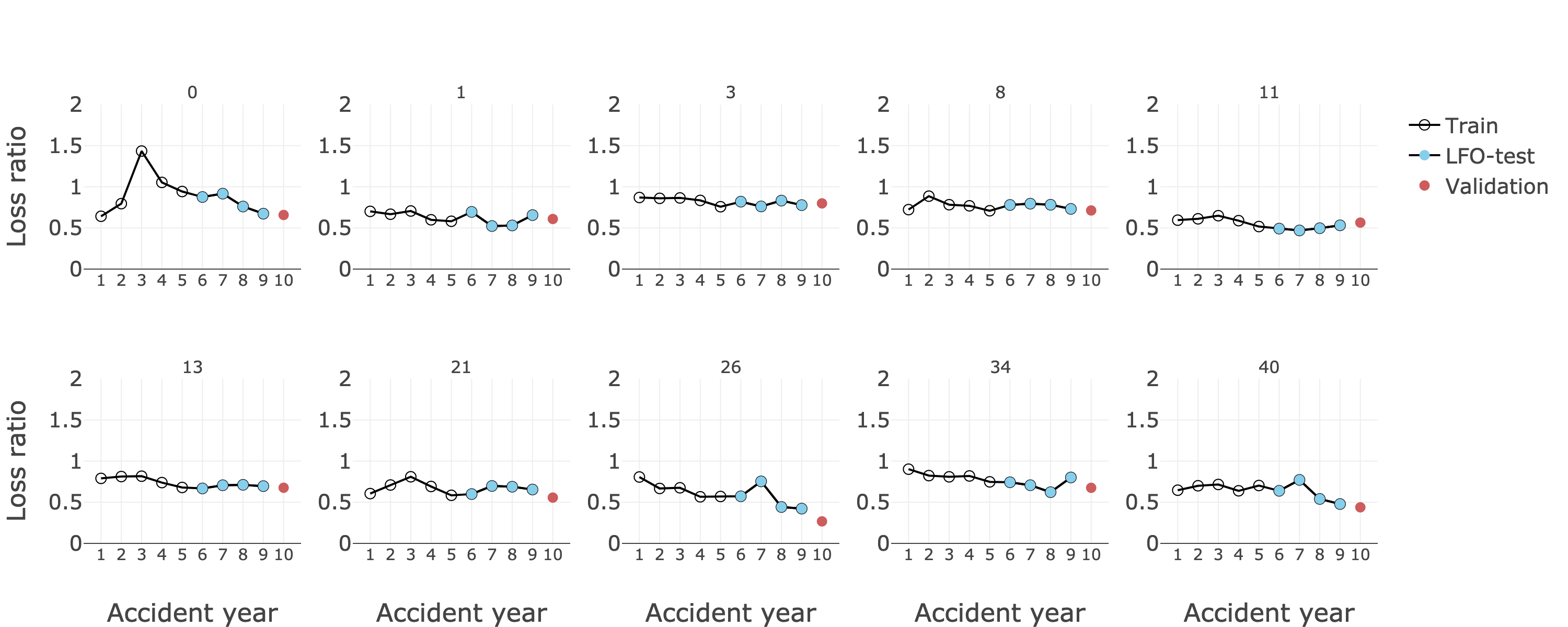}
    \caption{
        Ten randomly selected programs of 50 used for the forecasting example.
        Each program has 10 close-to-ultimate incurred loss ratios (circles), 
        developed to 10 years. Each candidate model is fit to all 50 programs through
        year 9 and predictions are validated at year 10 (red points). Leave-future-out
        cross-validation is used to predict years 6 through 9 (blue points) to create a 
        test-set of data to train the blending models.
    }
    \label{fig:forecast-programs}
 \end{figure}

We model incurred loss ratios (i.e. paid losses plus case reserves divided
by earned premium) for each of the 50 programs in the dataset as of development
lag 10 years (i.e. the final column in table \ref{tab:triangle-splits}). 
Incurred losses are used rather than paid losses to 
account for case reserves that better reflect ultimate losses, although both 
paid and incurred losses at 10 years of development are very similar.
With only 10 data points per program, we first fit our set of candidate
models through to year 9, and hold out year 10 for each program as a validation
point. To train the stacking models, we use one step-ahead leave-future-out (LFO) cross 
validation \citep{burkner2020lfo} on points 6 through 9, which involved
iteratively fitting each model through to point $m - 1$ to predict
point $m = (6, ..., 9)$ (see Figure \ref{fig:forecast-programs}).

We chose six candidate models of varying complexity. 
As in the loss development example above,
we used a gamma distribution for the likelihood
distribution, parameterized by its mean and variance, which are
converted to shape and inverse scale parameters before sampling in \texttt{CmdStan}
via \texttt{CmdStanPy}.
The Constant model represents the $i$th program's loss ratio, $LR$,
for the $m$th accident year by
constant mean and variance parameters, where programs' mean loss ratios are 
partially-pooled in a hierarchical structure:

\begin{align}
\begin{split}
    LR_{im} &\sim \mathrm{Gamma}(\mu_{im}, \sigma^2)\\
    \mu_{im} &= \exp(\alpha + u_{i})\\
    u_{i} &\sim \mathrm{Normal}(0, \sigma_{u})\\
    \sigma_{u} &\sim \mathrm{Normal}^{+}(0, 1)\\
    \sigma^2 &\sim \mathrm{Lognormal}(0, 1)\\
    \alpha &\sim \mathrm{Normal}(0, 1)
\end{split}
\end{align}

The Linear model assumes a linear regression of loss ratios, $LR$,
on standardized accident year, where program-level intercepts and
slopes are assumed correlated with covariance matrix $\Sigma$:

\begin{align}
\begin{split}
    LR_{im} &\sim \mathrm{Gamma}(\mu_{im}, \sigma^2)\\
    \mu_{im} &= \exp(\alpha + u_{i1} + (\beta + u_{i2}) z(\mathrm{AY}_{im}))\\
    (u_{1}, u_{2})' &\sim \mathrm{MultivariateNormal}((0, 0)', \Sigma)\\
    \Sigma &= \mathrm{diag}(\sigma_{u}) \Omega \mathrm{diag}(\sigma_u)\\
    \Omega &= \mathrm{LKJ}(2.0)\\
    \sigma_{u} &\sim \mathrm{Normal}^{+}(0, 1)\\
    \sigma^2 &\sim \mathrm{Lognormal}(0, 1)\\
\end{split}
\end{align}

where the function $z$ standardizes the data,
$\sigma_u = (\sigma_{u_{1}}, \sigma_{u_{2}})'$ is the vector of
standard deviations, and $\Omega$ is the correlation matrix, which
receives an LKJ prior distribution. In practice,
we estimate the Cholesky factor of the correlation matrix, $\Omega$, 
and use a non-centered parameterization to aid computational efficiency. 

The third model, Linear-Heteroskedastic (Linear-Hk), is the same as the linear model above, except for
the estimation of heteroskedastic variances across programs:

\begin{align}
\begin{split}
    LR_{im} &\sim \mathrm{Gamma}(\mu_{im}, \sigma_{i}^{2})\\
    \mu_{im} &= \exp(\alpha + u_{i1} + (\beta + u_{i2}) z(\mathrm{AY}_{im}))\\
    \sigma_{i}^{2} &= \exp(\delta + u_{i3})\\ 
    (u_{1}, u_{2}, u_{3})' &\sim \mathrm{MultivariateNormal}((0, 0, 0)', \Sigma)\\
\end{split}
\end{align}

The final three models are time series-specific, accounting for
correlation structure in the data. Model 4 is an autorgessive lag-1
(AR(1)) model with
drift (i.e. intercepts), which partially pools program-level
estimates in the long-term mean loss ratios, the autogressive
parameter, and the variances:

\begin{align}
\begin{split}
    LR_{im} &\sim \mathrm{Gamma}(\mu_{im}, \sigma_{i}^{2})\\
    \mu_{im} &= \exp((1 - \phi_{i}) (\alpha + u_{i1}) + \phi_{i} \log LR_{im-1})\\
    \phi_{i} &= \mathrm{logit}^{-1}(\phi^{*} + u_{i2}) \cdot 2 - 1\\
    \sigma_{i}^{2} &= \exp(\delta + u_{i3})\\ 
    (u_{1}, u_{2}, u_{3})' &\sim \mathrm{MultivariateNormal}((0, 0, 0)', \Sigma)\\
    \sigma_{u} &\sim \mathrm{Normal}^{+}(0, 1)\\
    \delta &\sim \mathrm{Normal}(0, 1)\\
    \alpha &\sim \mathrm{Normal}(0, 1)\\
    \phi^{*} &\sim \mathrm{Normal}(0, 1)
\end{split}
\end{align}

where $\phi_{i}$ is a transformed parameter from the unconstrained scale
of $\phi^{*} + u_{i2}$ such that $\phi_{i} \in (-1, 1)$. The first time
point is modeled separately for each program as initial values.

Model 5 is a state-space model (SSM) with an AR(1) structure
on the latent loss ratio level. The model is the same as model 4, but
now $\mu_{im}$ is a function of its lag-1 self, rather than the observed
loss ratios:

\begin{align}
\begin{split}
    \mu_{im} &= \exp((1 - \phi_{i}) (\alpha + u_{i1}) + \phi_{i} \mu_{im-1})\\
\end{split}
\end{align}

Finally, model 6 includes a Gaussian process model (GP) on accident
years for the mean function:

\begin{align}
\begin{split}
    LR_{im} &\sim \mathrm{Gamma}(\mu_{im}, \sigma_{i}^{2})\\
    \mu_{im} &= \exp(\gamma + u_{i1} + \mathrm{GP}(AY, \omega, \zeta))\\
    \sigma_{i}^{2} &= \exp(\delta + u_{i2})\\ 
    (u_{1}, u_{2})' &\sim \mathrm{MultivariateNormal}((0, 0)', \Sigma)\\
    \sigma_{u} &\sim \mathrm{Normal}^{+}(0, 1)\\
    \Sigma &= \mathrm{diag}(\sigma_{u}) \Omega \mathrm{diag}(\sigma_u)\\
    \Omega &= \mathrm{LKJ}(2.0)\\
    \omega &\sim \mathrm{Lognormal}(0, 0.25)\\
    \zeta &\sim \mathrm{Lognormal}(0, 0.5)\\
    \gamma &\sim \mathrm{Normal}(0, 1)\\
    \delta &\sim \mathrm{Normal}(0, 1)\\
\end{split}
\end{align}

The GP function uses the squared exponential quadratic kernel to define
the covariance matrix of the accident years, where $\omega$ and $\zeta$ 
are the standard deviation and length-scale parameters
of the kernel, respectively. As in the models above, we include intercepts
($\gamma + u_{i1}$) and heteroskedastic variances ($\sigma_{i}^{2}$).

Each model was fit to the data through accident year 9, and posterior 
predictives and log posterior predictive densities calculated on the validation set, 
year 10. Subsequently, the LFO cross validation
routine was run for each model from time points 6 to 9 to estimate out-of-sample
log posterior predictive densities, as depicted
in Figure \ref{fig:forecast-programs}. Lastly, we used the 200 LFO
log posterior density estimates (4 per 50 programs) to train our five
stacking models, and blend the posterior predictions from each
model according to the five sets of model weights. For the hierarchical stacking
model, we included a (standardized) continuous covariate of earned premium,
and discrete covariates of accident year and program. Earned premium is
included as a continuous covariate because, as the variance of
losses is expected to be proportional to premium, the variance of
loss ratios is expected to be inversely proportional to earned
premium.

\begin{figure}
    \centering
    \includegraphics[scale=0.17]{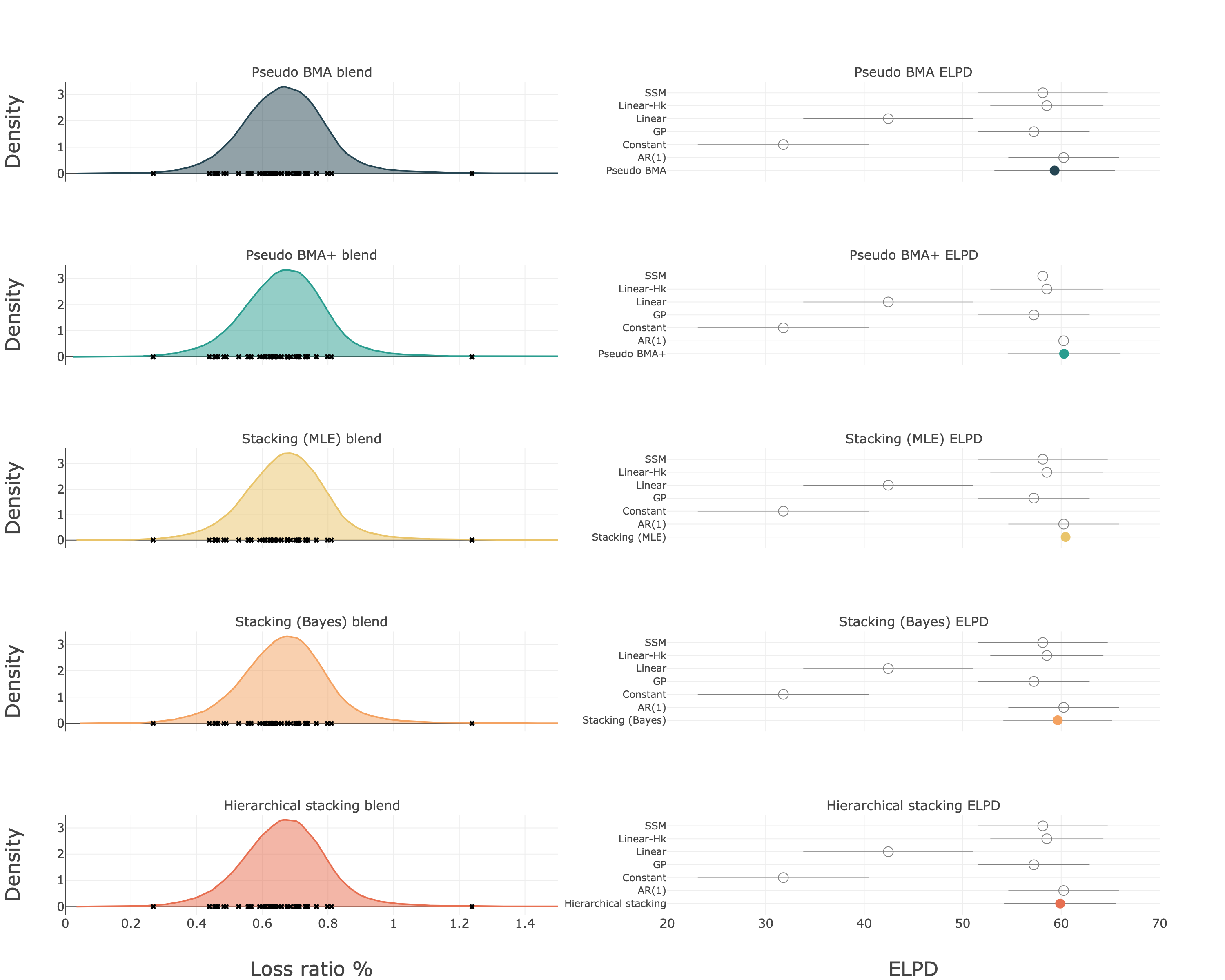}
    \caption{
        Blended posterior distributions and ELPD values from the
        forecasting example. In the first column,
        black crosses on the blended posterior
        densities indicate the real data. In the second column,
        ELPD point estimates and 1 standard deviation ranges are
        shown by open and filled circles, denoting candidate model
        and blended ELPD estimates, respectively, and the horizontal
        line segments. 
    }
    \label{fig:forecast-scores}
\end{figure}

\begin{figure}
    \centering
    \includegraphics[scale=0.13]{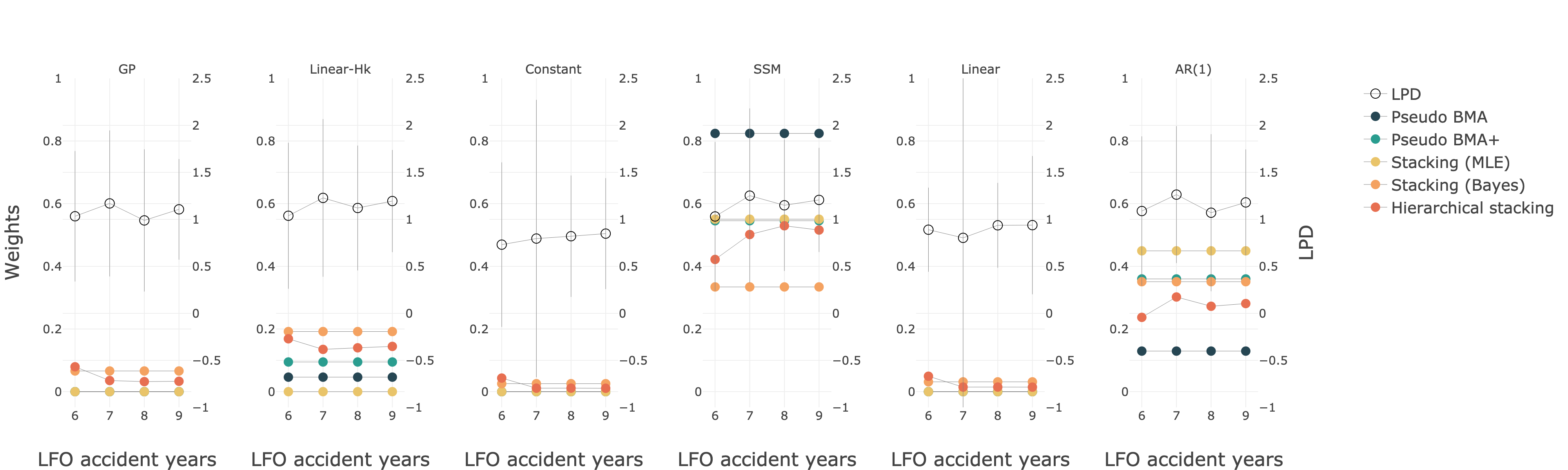}
    \caption{
        Model weights and log posterior predictive density (LPD) values from
        the forecasting model. Model weights are constant across leave-future-out
        (LFO) accident years for every model except the hierarchical stacking
        model, which included accident year as a discrete covariate. The left
        y-axis illustrates the weight values for each of the blending models.
        The right y-axis denotes the mean (open circles) and 1 standard deviation
        across programs (vertical line segments) LPD values.
    }
    \label{fig:forecast-weights}
\end{figure}

The blended posterior distributions and the ELPDs for each of the
six models and blended distributions is shown in Figure
\ref{fig:forecast-scores}. All blending models produce distributions
that performn better than most of the candidate models individually,
although only the pseudo-BMA+ and MLE stacking models beat the
best-performing candidate model, which is the AR(1) model in this
instance. Inspecting the model weights in Figure \ref{fig:forecast-weights},
the SSM model achieves the highest weight in the LFO test data set,
followed by the AR(1), Linear-Hk and GP models, respectively. The MLE stacking
and pseudo-BMA+ blends give most weight to the SSM and AR(1) models, and regularize
the other models' weights closer to zero than the remaining blends.
The Bayesian stacking and hierarchical stacking models give more weight to
Linear-Hk and GP models than pseudo-BMA+ and MLE stacking, while pseudo-BMA
gives significantly more weight to the SSM model than any other blend.
While the SSM model is a clear winner on the test data, the AR(1) model
performs best on the validation set. 
Although blending does not always produce the best predictive distribution
in this instance, the example highlights that performance on finite test data
will not always generalize to future data. When validation
data is not available, and we do not know the best performing candidate model,
blending can still produce a distribution that is
remarkably close to the best-performing candidate model on out-of-sample data.

\section{Conclusion}
\texttt{BayesBlend} is the first opensource software of its kind, allowing 
users to easily fit pseudo-BMA and stacking models to candidate models
of interest and blend together candidate model predictions to improve
predictive performance. Our real-world examples illustrate the utility
of model blending in an actuarial context, but pseudo-BMA and stacking
are techniques that can be applied to model predictions from any domain.
We encourage readers interested in learning more to refer to our GitHub 
repository at \url{https://github.com/LedgerInvesting/bayesblend}, where you can 
find links to further package documentation, documentation on how to 
contribute to \texttt{BayesBlend}, a discussion forum for questions related 
to \texttt{BayesBlend}, and issue tracking for new features that we have in 
the development pipeline. 

In the future, we plan to continue adding integrations to \texttt{BayesBlend}
to make it better interface with common Bayesian modeling libraries and 
probabilistic programming languages more generally. We also plan to add more
quality of life features, including plotting functionality for better model
inspection, model comparison functionality, and more. Additionally, although 
our focus has been on Bayesian candidate models, in principle it is possible
to use \texttt{BayesBlend} to blend together model predictions from non-Bayesian 
models. We encourage users to post on the GitHub dicussion forum if they are 
interested in features that are not currently supported.

\section{Competing interests}
We declare no competing interests.

\bibliographystyle{apa}
\bibliography{\src/references.bib}
\end{document}